\documentclass[twocolumn,journal]{IEEEtran}
\usepackage{graphicx}
\usepackage{amsmath}
\usepackage{amssymb}
\usepackage{comment}

\usepackage{amsthm}
\usepackage{cite}
\usepackage{float}
\usepackage{color}
\usepackage{balance}

\begin{document}
\title{\LARGE Cooperative Access Schemes for Efficient SWIPT \\ Transmissions in Cognitive Radio Networks}
\author{Ahmed El Shafie$^\dagger$, Naofal Al-Dhahir$^\dagger$ and Ridha Hamila$^*$\\
\begin{tabular}{c}
 $^\dagger$Electrical Engineering Dept., University of Texas at Dallas, USA. \\
  $^*$Electrical Engineering Dept., Qatar University, Doha, Qatar.
\end{tabular}
\thanks{This paper has been presented in globecom 2015.}
\thanks{This paper was made possible by NPRP grant number 6-070-2-024 from the Qatar National Research Fund (a member of Qatar Foundation). The statements made herein are solely the responsibility of the authors.}
}
\date{}
\maketitle
\thispagestyle{empty}
\pagestyle{empty}
\begin{abstract}
We investigate joint information and energy cooperative schemes in a slotted-time cognitive radio network with a primary transmitter-receiver pair and a set of secondary transmitter-receiver pairs. The primary transmitter is assumed to be an energy-harvesting node. We propose a three-stage cooperative transmission protocol. During the first stage, the primary user releases a portion of its time slot to the secondary nodes to send their data and to power the energy-harvesting primary transmitter from the secondary
radio-frequency signals. During the second stage, the primary transmitter sends its data to its destination and to the secondary nodes. During the third stage, the secondary nodes amplify and forward the primary data. We propose five different schemes for secondary access and powering the primary transmitter. We derive closed-form expressions for the primary and secondary rates for all the proposed schemes. Two of the proposed schemes use distributed beamforming to power the primary transmitter. We design a sparsity-aware relay-selection scheme based on the compressive sensing principles. Our numerical results demonstrate the gains of our proposed schemes for both the primary and secondary systems.
\end{abstract}
\begin{IEEEkeywords}
Cognitive radio, cooperation, energy harvesting, simultaneous wireless information and power transfer.
\end{IEEEkeywords}
\vspace{-0.1cm}
\section{Introduction}
\vspace{-0.0cm}
Energy-harvesting schemes extend the lifetime of wireless network nodes. Dynamic and efficient spectrum access can be achieved using cognitive radio techniques. The cognitive radio nodes dynamically and opportunistically access the primary-licensed frequency bands to enhance spectrum utilization efficiency. Recently, efficient integration of energy-harvesting
techniques into cognitive radio networks
has attracted a significant attention from both
industry and academic communities. In \cite{park2014achievable}, the authors analyze a theoretical upper-bound
on the maximum achievable throughput of an energy-harvesting secondary system. A new channel selection
criterion for an energy-harvesting secondary system is
developed in \cite{6814804}.

Since radio-frequency (RF) signals used for data transmission can also carry energy, simultaneous wireless information and power transfer (SWIPT) has received increased attention from researchers recently. In \cite{lee2013opportunistic}, the secondary transmitters (STs) harvest energy from the RF signals of the nearby active primary transmitters (PTs). Energy and information cooperation is proposed in \cite{6763046} where a primary user powers a secondary user which, in turn, relays the primary data. This paper studies joint information and energy cooperation where the PT sends information for
relaying and supplies the secondary system with energy as well.
In \cite{7032337}, the authors
propose an energy cooperation protocol where secondary users
cooperate with primary users to provide RF signals
for the primary users' energy harvesting and acquire spectrum opportunities. Both \cite{6763046,7032337} assumed that the PT releases a portion of its time slot for the secondary nodes.

Compressive sensing (CS) theory provides the conditions needed to reconstruct a long sparse vector from few noisy
measurements \cite{1580791,1614066} leading to significant applications in different areas including wireless sensor networks \cite{bajwa2006compressive}.
 The authors of \cite{fazel2011random} investigated a random-access CS-aided scheme
for underwater sensor networks. Their ultimate goal was to
design a power-efficient random data collection scheme.
A CS-aided medium access control (CS-MAC)
scheme is proposed in  \cite{lin2012compressive} where the access point (AP) allocates a random
sequence to each user.
All user requests for gaining uplink
transmissions are sent simultaneously in a
synchronous manner.
In \cite{naofal}, the authors proposed a new relay-selection technique based on sparse approximation theory for multiple antenna-relay selection with relay gain control.

In this paper, we propose a joint information and energy cooperative schemes for slotted-time cognitive radio networks. The contributions of this paper are summarized as follows. We design efficient schemes for energy and data \emph{transfer} using energy-harvesting techniques and sparse-approximation principles. We propose a three-stage scheme where a time slot is divided into three time intervals. In the first stage, the secondary nodes utilize a portion of the time slot to simultaneously send their data and power the PT. In the second stage, the PT transmits its data to its destination and the secondary nodes. In the third stage, the secondary nodes amplify and forward the received primary data. We propose \emph{five} different schemes for secondary spectrum access and PT powering.
      We propose a low-complexity multiple relay selection with gain control scheme based on CS principles to obtain a sparse set of relays such that the least square error of the primary data at its respective destination is minimized.



\begin{figure}
	\centering
		  \includegraphics[width=0.7\columnwidth, height=0.6\columnwidth]{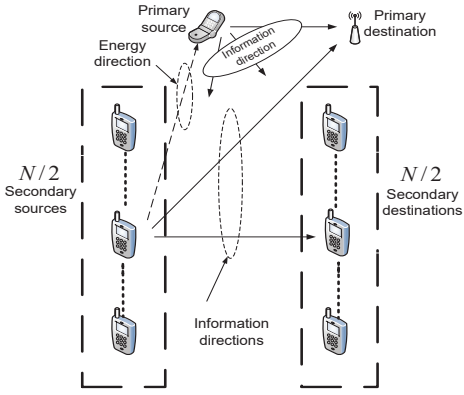}
	\caption{System model: the solid arrows represent the information transfer direction, while the dotted arrows represent the power transfer direction.}
	\label{fig00}
\vspace{-0.5cm}
\end{figure}
\vspace{-0.2cm}
\section{System Model and Assumptions}
\vspace{-0.1cm}
We consider a cognitive radio setting composed of a  primary transmitter-receiver pair and a set of $N/2$ secondary transmitter-receiver pairs as shown in Fig. \ref{fig00}. Thus, the total number of secondary nodes (i.e. transmitters and receivers) is $N$, where $N$ is assumed to be an even integer. The secondary nodes are labeled as $1,2,\dots, N$ where the transmitters are labeled as $1,2,\dots, N/2$ and the receivers are labeled as $N/2+1,N/2+2,\dots, N$. The respective receiver of transmitter $1 \le m \le N/2$ is node $N/2+1\le m+N/2\le N$. The $m$-th secondary node is denoted by ${\rm s}_m$. All nodes are equipped with a single antenna. The PT has RF
energy transfer capabilities and can harvest energy from the received
electromagnetic radiations.
We propose a three-stage scheme for performing the following two tasks: (1) simultaneous secondary data transmission and PT powering, (2) transmitting the primary data \emph{reliably} using cooperative relaying. The PT remains idle during the first stage to harvest energy from the ambient secondary RF transmissions. In the second stage, the PT transmits its data to the primary destination and a set of secondary relays that operate in the amplify-and-forward (AF) relaying mode. In the third stage, the secondary nodes forward the received signals from the primary transmission to the primary destination which combines the received signals using the maximal ratio combining (MRC) technique.

\begin{figure}
	\centering
		  \includegraphics[width=0.88\columnwidth]{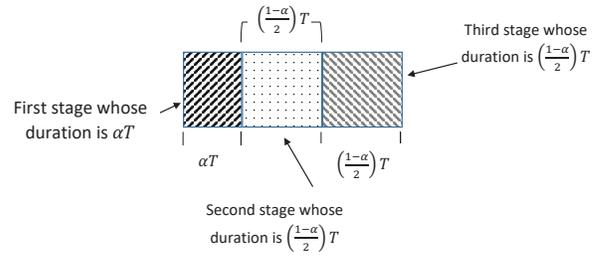}
	\caption{Time slot structure. The second and third stages have equal durations to enable the AF-relaying mode.}
	\label{fig0}
\vspace{-0.4cm}
\end{figure}

We assume a slotted-time system where the time is partitioned into slots each with a duration of $T$ time units. The PT occupies a bandwidth of $W$ Hz. The duration of the first stage is $0\le \alpha T\le T$ time units, whereas the durations of the second and the third stages are $\frac{1-\alpha}{2} T$ time units. The time slot structure is depicted in Fig. \ref{fig0}. A list of the used key variables is given in Table \ref{table1}.

Each link experiences a quasi-static block fading where the fading coefficient is assumed to be fixed during a time slot, but it changes from one time slot to another identically and independently. The channel coefficient from node $n_1$ to node $n_2$ is denoted by $h_{n_1,n_2}$. The channel gain, which is the squared magnitude of the channel coefficient, is denoted by $\theta_{n_1,n_2}=|h_{n_1,n_2}|^2$, where $|\cdot|$ denotes the absolute value.\footnote{Throughout this paper, we omit the time index from the symbols for notation convenience. However, we use it explicitly on several parameters to show the dependency of these parameters on current and previous time slots.} It is assumed
that all the channel coefficients are known to all the nodes through feedback or channel estimation \cite{naofal,4543070,6763046,7032337}. The thermal noise is modeled as an additive white Gaussian noise (AWGN) random process with zero mean and variance $\kappa$. We assume that the RF energy collected at the PT is converted to a direct current (DC) electricity with efficiency $0\le \eta \le 1$ which is a function
of the rectification process as well as the energy-harvesting
circuity. Furthermore, the processing energy consumed
by the PT's transmit circuitry is negligible \cite{6763046,7032337}. We assume that the PT has a constant energy supply of $E_{\rm p}$ energy units \cite{6763046,7032337}. In addition, we assume that the energy storage at the PT is unlimited \cite{6763046,7032337}. Each secondary user transmits its data with an average power of $P_{\rm s}$ Watts/Hz. Furthermore, we assume that the average power used for either powering the PT during the first stage or retransmitting its data during the third stage is $P_{\rm c}$ Watts/Hz, where we assume for simplicity of presentation that $P_{\rm c}\!=\!P_{\rm s}$.

In our proposed schemes, we assume that the secondary destinations can also be used to energize the PT and for relaying the primary data. This leads to a significant gain since we have additional $N/2$ potential nodes for cooperation. In the following sections, we discuss the system operation during each of the three proposed stages.
\vspace{-0.3cm}
\section{Proposed Secondary SWIPT Access Schemes}\label{sec3}
\vspace{-0.05cm}
 In the first stage, the STs occupy the spectrum and the primary transmitter
harvests the RF energy from their RF transmissions. The energy harvested is denoted by $E_{h,1}^t$. During this stage, we propose several policies, which differ in terms of implementation complexity, for the STs access and the primary powering processes.
\vspace{-0.4cm}
\subsection{First Proposed Secondary Access (First-PSA) Scheme}
\vspace{-0.1cm}
In this scheme, all secondary users share the available time for powering the PT and for sending their data. The STs access the channel \emph{one at a time}. We assume here that the secondary destinations do not participate in PT powering. Since we have $N/2$ STs, each ST is assigned only $\alpha/(N/2)$ of the time assigned for the first stage. Consequently, the energy harvested at the PT in joules/Hz is
\begin{equation} \small
E_{h,1}^t=P_{\rm s} \frac{\alpha}{N/2} T \eta \sum_{i=1}^{{N}/{2}} \theta_{{\rm s}_i,{\rm p}},
\end{equation}
where $\eta$ is the energy conversion efficiency. The expression of $E_{h,1}^t$ is explained as follows. Since one ST transmits over a duration of $T\alpha/(N/2)$ with average transmit power $P_{\rm s}$, the received energy at the PT during the $i$-th ST's transmission is $T\alpha/(N/2) P_{\rm s} \theta_{{\rm s}_i,{\rm p}}$. The summation over all ST's transmissions is $P_{\rm s} \frac{\alpha}{N/2} T \eta \sum_{i=1}^{N/2} \theta_{{\rm s}_i,{\rm p}}$.

Since an ST uses the channel for $T \alpha/(N/2)$ time units, the rate of the $i$-th ST in bits/sec/Hz is
\begin{equation} \small
\label{eqn2}
\mathcal{R}_i=\frac{\alpha}{N/2} \log_2\left( 1+\frac{P_{\rm s} \theta_{{\rm s}_i,{\rm s}_{i+N/2}}}{\kappa}\right).
\end{equation}
\begin{table}
\renewcommand{\arraystretch}{1}
\begin{center}
\begin{tabular}{ c |l  }
    \hline\hline
    Symbol & Description \\[5pt]\hline
    $T$ and $W$ & {\footnotesize Primary slot duration and channel bandwidth} \\[5pt]\hline
        $\alpha T$ & {\footnotesize Time duration assigned for the first stage} \\[5pt]\hline
        $\frac{1-\alpha}{2}T$ & {\footnotesize Time duration assigned for the second and third stages} \\[5pt]\hline
         $h_{{n_1},{n_2}}$ & {\footnotesize Channel coefficient between node $n_1$ and node $n_2$} \\[5pt]\hline
        $\theta_{n_1,n_2}$ & {\footnotesize Channel gain of the $n_1-n_2$ link} \\[5pt]\hline
  $N$ & {\footnotesize Number of secondary nodes} \\[5pt]\hline
    $P_{\rm s}$ & {\footnotesize Transmit power of a secondary node} \\[5pt]\hline
        $P_{\rm p}$ & {\footnotesize Transmit power of the PT} \\[5pt]\hline
           $E_{\rm p}$ & {\footnotesize Constant energy supply of the PT} \\[5pt]\hline
               & {\footnotesize Average energy harvested at the PT during the first stage}\\ $E^t_{h,1}$ & { \footnotesize of time slot $t$} \\[5pt]\hline
               & {\footnotesize Average energy harvested at the PT during the third stage}\\ $E^t_{h,3}$ & { \footnotesize  of time slot $t$} \\[5pt]\hline
     $\kappa$ & {\footnotesize Variance of the AWGN at a receiving node} \\[5pt]\hline
       $\eta$ & {\footnotesize RF-to-DC conversion efficiency} \\[5pt]\hline
        $g_n$ & {\footnotesize The gain associated with the $n$-th selected secondary relay} \\[5pt]\hline
\end{tabular}
\end{center}
\caption{List of Key Variables.}
\label{table1}
\vspace{-0.8cm}
\end{table}
\vspace{-0.8cm}
\subsection{Second Proposed Secondary Access (Second-PSA) Scheme}
\vspace{-0.00cm}
In this scheme, we propose that the STs transmit their data simultaneously over the whole time duration released by the primary user. That is, each secondary transmission takes place over the first $\alpha$ portion of the time slot. Thus, the energy collected at the PT due to the ambient RF transmissions is
\begin{equation} \small
E_{h,1}^t=P_{\rm s} \alpha T \eta \sum_{i=1}^{N/2} \theta_{{\rm s}_i,{\rm p}}.
\end{equation}
In this scheme, the secondary destinations cannot be used for powering the PT since they are busy decoding the data sent from the STs. The rate of the $i$-th ST is given by
\begin{equation} \small
\mathcal{R}_i={\alpha} \log_2\left( 1+\frac{P_{\rm s} \theta_{{\rm s}_i,{\rm s}_{i+N/2}}}{\kappa+ \sum_{\substack{{j=1}\\{j\ne i}}}^{N/2} P_{\rm s} \theta_{{\rm s}_j,{\rm s}_{j+N/2}}}\right).
\end{equation}
We note that there is interference among the secondary nodes such that each destination receives the data of its respective transmitter corrupted by both AWGN and the interference from $N/2-1$ other transmissions.
\vspace{-0.2cm}
\subsection{Third Proposed Secondary Access (Third-PSA) Scheme}
In this scheme, we assume that the ST having the maximum link gain among all STs to their destinations is selected for data transmission. Thus, this ST and its respective destination will communicate with each other during the first stage. At the same time, the secondary node from the remaining secondary nodes which has the maximum link gain among all secondary nodes to the PT will be selected for powering the PT. Note that the secondary node selected for powering the PT uses a known time-invariant signal which can be eliminated by the secondary destination before data decoding. Assuming that, in the given time slot, the $k$-th ST is the one which has the maximum gain to its destination, its rate is then given by
\begin{equation} \small
\label{eqn5}
\mathcal{R}_k={\alpha} \log_2\left( 1+\frac{{P_{\rm s}} \theta_{{\rm s}_k,{\rm s}_{k+N/2}}}{\kappa}\right).
\end{equation}
We emphasize that there is no interference at the destination because it knows the transmit signal by the node that powers the PT and also knows the channel gains of all nodes. Hence, it subtracts the interfering signal from the received signal.

Assuming that the $r$-th secondary node is the one which has the maximum gain to the PT, the energy harvested at the PT during this stage is given by \begin{equation} \small
E_{h,1}^t=P_{\rm s} \alpha T \eta \left(\theta_{{\rm s}_k,\rm p}+\theta_{{\rm s}_r,\rm p}\right).
\end{equation}
Note that portion of the harvested energy at the PT is due to the ST scheduled for data transmission, i.e., node ${\rm s}_{k}$.
\vspace{-0.1cm}
\subsection{Fourth Proposed Secondary Access (Fourth-PSA) Scheme}
In this scheme, we propose to select the user with the maximum link to its destination for data transmission while using a beamformer via selecting $K\ge2$ of the remaining secondary nodes to maximize the received energy at the PT while completely eliminating the interference at the receiver of the secondary node scheduled for data transmission. This can be simply done by assuming that nodes power (energize) the PT using a known but time-invariant signal to all nodes in the network including all destinations. In this case, the secondary node scheduled for data transmission and the nodes selected for powering the PT occupy the whole $\alpha T$ seconds. Assuming that the $k$-th ST is the one which has the maximum link gain to its destination, the energy harvested at the PT during the first stage is
\begin{equation} \small
E_{h,1}^t=P_{\rm s} \alpha T \eta \left(\theta_{{\rm s}_k,\rm p}+\left|\sum_{\substack{{j\in \Omega}\\{k,k+N/2\notin \Omega}}} \beta_j^\dagger h_{{\rm s}_j,{\rm p}}\right|^2\right),
\end{equation}
where $\Omega$ with cardinality $K$ is the set of nodes selected for powering the PT and $\beta^\dagger_j=h^\dagger_{{\rm s}_j,{\rm p}}/\sqrt{\sum_{j\in \Omega} |h_{{\rm s}_j,{\rm p}}|^2}$ is the complex conjugate of the weight $\beta_j$ used at the $j$-th secondary node belonging to the set $\Omega$. Note that $k$ and its respective destination are not in $\Omega$.
%
The rate of the $k$-th ST is
\begin{equation} \small
\label{eqn8}
\mathcal{R}_k={\alpha} \log_2\left( 1+\frac{{P_s} \theta_{{\rm s}_k,{\rm s}_{k+N/2}}}{\kappa}\right).
\end{equation}
\vspace{-0.2cm}
\subsection{Fifth Proposed Secondary Access (Fifth-PSA) Scheme}
In this scheme, the time duration of the first stage is divided equally among the STs for data transmissions. When a node is selected for transmission, all other nodes cooperatively energize the PT using the appropriate beamformer. We assume that only $K\ge 2$ of the remaining secondary nodes are used to create the beamformer to maximize the received energy at the PT while completely eliminating the interference at the receiver of the ST scheduled for data transmission.
 The energy harvested at the PT during the first stage is given by
\begin{equation} \small
E_{h,1}^t=P_{\rm s} \frac{\alpha}{N/2} T \eta \sum_{i=1}^{N/2} \left(\theta_{{\rm s}_i,\rm p}+\left|\sum_{\substack{{j\in \Omega}\\{i,i+N/2\notin \Omega}}} \beta_j^\dagger h_{{\rm s}_j,{\rm p}}\right|^2\right).
\end{equation}

The rate of the $i$-th ST is given by
\begin{equation} \small
\label{eqn10}
\mathcal{R}_i=\frac{\alpha}{N/2} \log_2\left( 1+\frac{P_{\rm s}  \theta_{{\rm s}_i,{\rm s}_{i+N/2}}}{\kappa}\right).
\end{equation}

We conclude this section by mentioning that the rates of the STs under the first and fifth PSA schemes (c.f. Equations (\ref{eqn2}) and (\ref{eqn10})) and the third and fourth PSA schemes (c.f. Equations (\ref{eqn5}) and (\ref{eqn8})) are equal, respectively.


\section{Primary Data Transmission}\label{sec4}
\subsection{Cooperative Relaying}
In the second stage, the secondary nodes cease their transmissions and
the PT transmits with average power $P_{\rm p}^t$ in the
\emph{first half} of the remaining $(1-\alpha)$ portion of the time slot. The PT's average transmit power in Watts/Hz is given by
\begin{equation} \small
\label{polop}
P_{\rm p}^{t}=2\frac{E^{t}_{\rm p}+E_{h,1}^{t}+E_{h,2}^{t-1}}{(1-\alpha)T},
\end{equation}
where the factor of $2$ in (\ref{polop}) is due to the fact that only $(1-\alpha)/2$ of the remaining time in the time slot is used during this stage, and $E_{h,2}^{t-1}$ is the amount of energy harvested during the third stage of the previous time slot as will be discussed later.
The received signal from the PT to the primary destination is
 \begin{equation} \small
\small \begin{split} \small
y^{(2)}_{\rm pd}=h_{\rm p,pd} x  + w_2,
     \end{split}
\end{equation}
where $x$ is the transmitted signal
from the PT with average transmission power $P_{\rm p}^t$,
the superscript of $y_{\rm pd}^{(\cdot)}$ represents the stage number, and $w_2$ is the noise at the primary destination during the second stage. The received signal at the secondary relays is given by
 \begin{equation} \small
\small \begin{split} \small
\mathbf{y}_S=\mathbf{H}_{{\rm p},S} \ x+\mathbf{v},
     \end{split}
\end{equation}
where $\mathbf{v}=[v_1,v_2,\dots,v_n,\dots,v_N]^\mathbb{T}$ is the AWGN vector whose $n$-th
component represents the AWGN at the $n$-th secondary node, $\mathbf{H}_{{\rm p},S}=[h_{{\rm p},{\rm s}_1},h_{{\rm p},{\rm s}_2},\dots,h_{{\rm p},{\rm s}_N}]^\mathbb{T}$ and $(\cdot)^\mathbb{T}$ denotes the transpose.
\vspace{-0.25cm}
\subsection{Sparse Relay Selection}
In the third stage, the secondary nodes amplify and forward the primary data.
We propose a multiple relay-selection scheme
for the relay network at hand where
the PT transmits its data to its destination and the secondary nodes. First, we
formulate the problem of minimum mean square error (MMSE) multiple secondary relay selection with gain control. We define the gain vector $\mathbf{g}=[g_1,g_2,\dots,g_n,\dots,g_N]^\mathbb{T}$ whose $n$-th
element $g_n$ is the gain associated with the $n$-th selected secondary relay. The received signal at the primary destination from the relays' transmissions is given by
 \begin{equation} \small
\small \begin{split} \small
y^{(3)}_{\rm pd}=\mathbf{g}^\dagger \boldsymbol{h} x+\mathbf{g}^\dagger \tilde{\mathbf{v}} + w_3,
     \end{split}
\end{equation}
where $\boldsymbol{h}\!=\![h_{{\rm s}_1,{\rm pd}} h_{{\rm p},{\rm s}_1}, \dots, h_{{\rm s}_N,{\rm pd}} h_{{\rm p},{\rm s}_N}]^\mathbb{T}$, $\tilde{\mathbf{v}}=[h_{{\rm s}_1,{\rm pd}} v_1, \dots, h_{{\rm s}_N,{\rm pd}} v_N]^\mathbb{T}$ are the relayed noise elements associated with the $n$-th secondary node, and $w_3$ is the noise signal at the primary destination during the third stage.

Defining a relay-selection vector $\mathbf{g}$,
the error signal is defined as follows
 \begin{equation} \small
\small \begin{split} \small
e=\left|x-\left(\mathbf{g}^\dagger \boldsymbol{h} x+\mathbf{g}^\dagger \tilde{\mathbf{v}} +  w_3\right)\right|^2.
     \end{split}
\end{equation}
Hence, the MSE at the primary destination can be written as
 \begin{equation} \small
\small \begin{split} \small
\mathbb{E}[e]&=P^t_{\rm p}  -\mathbf{g}^\dagger\boldsymbol{h}P^t_{\rm p}-\boldsymbol{h}^\dagger P^t_{\rm p}\mathbf{g} +\mathbf{g}^\dagger(\sqrt{P_{\rm p}^t} \boldsymbol{h}\boldsymbol{h}^\dagger+\boldsymbol{R}_{\tilde{\mathbf{v}} \tilde{\mathbf{v}}})\mathbf{g}+\kappa\\& =P^t_{\rm p} \!-\!\mathbf{g}^\dagger \tilde{ \boldsymbol{h}}-\tilde{ \boldsymbol{h}}^\dagger \mathbf{g}+\mathbf{g}^\dagger  \boldsymbol{R} \mathbf{g}+\kappa,
     \end{split}
\end{equation}
where $\mathbb{E}[\cdot]$ is the expected value operator, $\boldsymbol{R}~=~\sqrt{P_{\rm p}^t} \boldsymbol{h}\boldsymbol{h}^\dagger~+~\boldsymbol{R}_{\tilde{\mathbf{v}} \tilde{\mathbf{v}}}$, $\tilde{\boldsymbol{h}}=\boldsymbol{h}P^t_{\rm p}$, and the covariance matrix of the relayed noise vector is $\boldsymbol{R}_{\tilde{\mathbf{v}} \tilde{\mathbf{v}}} =
\mathbb{E}[\tilde{\mathbf{v}} \tilde{\mathbf{v}}^\dagger] ={\rm diag}(\sigma_{\tilde{\mathbf{v}}_1},\sigma_{\tilde{\mathbf{v}}_2},\dots, \sigma_{\tilde{\mathbf{v}}_n},\dots, \sigma_{\tilde{\mathbf{v}}_N})$ whose $n$-th element is
$\sigma_{\tilde{\mathbf{v}}_n}^2=|h_{{\rm p},{\rm s}_n}|^2 \kappa=\theta_{{\rm p},{\rm s}_n} \kappa$, where ${\rm diag}(\cdot)$ denotes a diagonal matrix with the enclosed elements as its diagonal entries. Define the Cholesky factorization, $\boldsymbol{R} = L L^\dagger$ where $L$ is an $N \times N$
lower-triangular matrix. Then, the MSE can be rewritten as follows
 \begin{equation} \small
\small \begin{split} \small
{\rm MSE}=P^t_{\rm p}\!-\!\mathbf{g}^\dagger L L^{-1} \tilde{ \boldsymbol{h}}\!-\!\tilde{ \boldsymbol{h}}^\dagger {L^{-1}}^\dagger L^\dagger \mathbf{g}\!+\!\mathbf{g}^\dagger  L L^\dagger \mathbf{g}+\kappa.
     \end{split}
\end{equation}
By completing the square, we can write
 \begin{equation} \small
\small \begin{split} \small
{\rm MSE}=\underbrace{P^t_{\rm p}-\tilde{ \boldsymbol{h}}^\dagger {L^{-1}}^\dagger L^\dagger +\kappa}_{{\rm MSE}_{\min}}+ \underbrace{\| L^\dagger \mathbf{g}-L^{-1} \tilde{\boldsymbol{h}}\|_2^2}_{{\rm MSE}_{\rm excess}},
     \end{split}
\end{equation}
where $\|\cdot\|_2$ denotes the $\ell_2$-norm. We note that the MSE is decomposed into two quantities. The first quantity $\rm MSE_{\min}$ does not depend on $\mathbf{g}$, while the second quantity ${{\rm MSE}_{\rm excess}}$ depends on $\mathbf{g}$. Hence, the MSE is minimized
by minimizing ${{\rm MSE}_{\rm excess}}$ which can be
tuned through the relay gain vector $\mathbf{g}$. The optimal weight vector that minimizes the MSE, denoted by $\mathbf{g}^\star$, is given by
 \begin{equation} \small
\small \begin{split} \small
\mathbf{g}^\star={L^{-1}}^\dagger L^{-1} \tilde{\boldsymbol{h}}= \boldsymbol{R}^{-1} \tilde{\boldsymbol{h}}.
     \end{split}
\end{equation}
In general, $\mathbf{g}^\star$ is not sparse and, hence, the complexity of
computing and implementing it is proportionally increasing
with $N$ which can be large. Any choice for
$\mathbf{g}$ different from $\mathbf{g}^\star$ increases ${{\rm MSE}_{\rm excess}}$ which translates into
performance degradation. A practical performance-complexity
trade-off can be achieved if we design a sparse $\mathbf{g}$ such that
${{\rm MSE}_{\rm excess}}\le \epsilon$  where  $\epsilon> 0$ controls the tolerable performance
loss in terms of MSE increase.
To select multiple
relays which minimize MSE, there are two classes of sparse approximation algorithms: convex optimization and greedy algorithms. The greedy algorithms are more suitable to the relay-selection problem proposed in this paper due to its low complexity. The orthogonal matching pursuit (OMP)
algorithm in \cite{tropp2007signal} is used for recovery of the sparse vector $\mathbf{g}$.
It takes the measurement vector $\mathbf{y}$ whose size is $M\times 1$, measurement matrix $\mathcal{A}$ whose size is  $M \times N$, and a certain stopping criterion as its inputs
and computes an $N$-dimensional sparse solution $\tilde{\mathbf{x}}$ for the unknown vector
$\mathbf{x}$ as its output. Hence, we denote the OMP operation by
$\tilde {\mathbf{x}} = {\rm OMP} (\mathbf{y}, \mathcal{A},{\rm stopping \ criterion})$. The stopping criterion
can be a predefined sparsity level (number of nonzero entries) of $\mathbf{x}$ or an upperbound on the norm of the residual error
term  $\|\mathbf{y}-\mathcal{A}\mathbf{x}\|^2_2$.

%

In our relay network formulation, the OMP algorithm attempts to find, in
each iteration, one column of the matrix $L^\dagger$ which is the
most correlated with the residual error vector obtained by
subtracting the contributions of the selected secondary relays in the
previous iteration from the vector $L^{-1} \tilde{\boldsymbol{h}}$. We consider the
OMP algorithm with the number of nonzero elements as our stopping criterion. That is, the sparse gain control vector is
\begin{equation} \small
\mathbf{g}_{\rm omp}={\rm OMP}\left( L^\dagger,L^{-1} \tilde{\boldsymbol{h}}, K_R \right),
\end{equation}
where $K_R$ is the total number of selected relays. Since the average transmit power by the relays is fixed, we have
 \begin{equation} \small
\small \begin{split} \small
\label{plplp}
\mathbb{E}\left[ \mathbf{g}^\dagger \mathbf{y}_S {\mathbf{y}_S}^\dagger \mathbf{g}\right] &=\mathbf{g}^\dagger  \mathbb{E}\left[( \mathbf{H}_{{\rm p},S} x+\mathbf{v})(\mathbf{H}_{{\rm p},S} x+\mathbf{v})^\dagger\right]\mathbf{g}\\& =\mathbf{g}^\dagger  \left( P^t_{\rm p} \mathbf{H}_{{\rm p},S}  {\mathbf{H}_{{\rm p},S}}^\dagger+ \kappa \mathbf{I}_N\right)\mathbf{g}=P_{\rm s},
     \end{split}
\end{equation}
where $P_{\rm s}\ge 0$ is the average total power constraint, and $\mathbf{I}_N$ is the identity matrix with size $N\times N$. Typically, the output of the OMP, which is a sparse control vector that minimizes the MSE, does not satisfy the power constraint in (\ref{plplp}). To satisfy this constraint, the output of the OMP must be multiplied by the following factor
 \begin{equation} \small
\small \begin{split} \small
\varrho=\sqrt{\frac{P_{\rm s}}{{\mathbf{g}_{\rm omp}}^\dagger  \left( P^t_{\rm p} \mathbf{H}_{{\rm p},S}  {\mathbf{H}_{{\rm p},S}}^\dagger+ \kappa \mathbf{I}_N\right)\mathbf{g}_{\rm omp}}}.
     \end{split}
\end{equation}
Hence, the optimal sparse gain control vector becomes
 \begin{equation} \small
\small \begin{split} \small
\mathbf{g}^*&=\mathbf{g}_{\rm omp} \ \varrho=\mathbf{g}_{\rm omp}\sqrt{\frac{P_{\rm s}}{{\mathbf{g}_{\rm omp}}^\dagger  \left( P^t_{\rm p} \mathbf{H}_{{\rm p},S}  {\mathbf{H}_{{\rm p},S}}^\dagger+ \kappa \mathbf{I}_N\right)\mathbf{g}_{\rm omp}}}.
     \end{split}
\end{equation}

{\bf{\underline{Remark}}}:\emph{ By constraining the relay gain vector to be sparse, our relay-selection approach achieves the minimum MSE within a tolerable value $\epsilon$, while reducing the number of selected relays.
This, in turn, reduces the implementation complexity due to AF protocol
signaling overhead and power consumption in the RF front-ends
at the secondary relays. It should be mentioned that AF relays must buffer the received
signals in the second stage of the time slot until they are amplified-then-transmitted in the third stage.  This buffering operation of data is efficiently done digitally
at baseband rather than in analog domain. Hence, there is a need for power-consuming down/up
conversion operations even for AF relays~\cite{naofal}.}

We note that the PT remains silent during the third stage, hence it can harvest more energy from the secondary RF transmissions which are used to forward the amplified primary data.
%
The received signal at the PT due to the secondary relays' transmissions is given by
 \begin{equation} \small
\small \begin{split} \small
y_{\rm p}=\mathbf{g}^\dagger \boldsymbol{h}_{\rm p} x+\mathbf{g}^\dagger \tilde{\mathbf{v}}_{\rm p} + w_{\rm p},
     \end{split}
\end{equation}
where $w_{\rm p}$ is the noise value at the PT during the third stage, and $\boldsymbol{h}_{\rm p}=[h_{{\rm s}_1,{\rm p}} h_{{\rm p},{\rm s}_1}, \dots,h_{{\rm s}_N,{\rm p}} h_{{\rm p},{\rm s}_N}]^\mathbb{T}$ and $\tilde{\mathbf{v}}_{\rm p}=[h_{{\rm s}_1,{\rm p}} v_1, \dots, h_{{\rm s}_N,{\rm p}} v_N]^\mathbb{T}$ are the relayed noise
elements associated with the $n$-th secondary node, respectively. The expected value of $y_{\rm p} y_{\rm p}^\dagger$ is given by
 \begin{equation} \small
\small \begin{split} \small
\mathbb{E}[y_{\rm p}y_{\rm p}^\dagger]=\mathbf{g}^\dagger \boldsymbol{h}_{\rm p} \boldsymbol{h}_{\rm p}^\dagger \mathbf{g}  P^t_{\rm p}+\mathbf{g}^\dagger \boldsymbol{R}_{\tilde{\mathbf{v}}_{\rm p} \tilde{\mathbf{v}}_{\rm p}} \mathbf{g} + \kappa,
     \end{split}
\end{equation}
where $P^t_{\rm p}$ is the power of the data transmitted by the PT during time slot $t$, and $\boldsymbol{R}_{\tilde{\mathbf{v}}_{\rm p} \tilde{\mathbf{v}}_{\rm p}} =
\mathbb{E}[\tilde{\mathbf{v}}_{\rm p} \tilde{\mathbf{v}}_{\rm p}^\dagger] ={\rm diag}(\sigma_{\tilde{\mathbf{v}}_{\rm p,1}},\sigma_{\tilde{\mathbf{v}}_{\rm p,2}},\dots, \sigma_{\tilde{\mathbf{v}}_{{\rm p},n}},\dots, \sigma_{\tilde{\mathbf{v}}_{{\rm p},N}})$ whose $n$-th element is
$\sigma_{\tilde{\mathbf{v}}_{{\rm p},n}}^2=|h_{{\rm s}_n,{\rm p}}|^2 \kappa=\theta_{{\rm s}_n,{\rm p}} \kappa$. Due to channel reciprocity, we have $h_{i,j}=h_{j,i}$; hence, $\boldsymbol{R}_{\tilde{\mathbf{v}}_{\rm p} \tilde{\mathbf{v}}_{\rm p}} =\boldsymbol{R}_{\tilde{\mathbf{v}} \tilde{\mathbf{v}}} $. We note here that the power used by the PT during the second stage of the time slot will be amplified-and-forwarded by the relays and a portion of it will be re-harvested by the PT. The energy harvested at the PT during the third stage is given by
\begin{equation} \small
E^t_{h,2}=\left(\mathbf{g}^\dagger \boldsymbol{h}_{\rm p} \boldsymbol{h}_{\rm p}^\dagger \mathbf{g}  P^t_{\rm p}+\mathbf{g}^\dagger \boldsymbol{R}_{\tilde{\mathbf{v}}_{\rm p} \tilde{\mathbf{v}}_{\rm p}} \mathbf{g}\right) \frac{1-\alpha}{2} T \eta,
\end{equation}
where $\mathbf{g}^\dagger \boldsymbol{h}_{\rm p} \boldsymbol{h}_{\rm p}^\dagger \mathbf{g}  P^t_{\rm p}+\mathbf{g}^\dagger \boldsymbol{R}_{\tilde{\mathbf{v}}_{\rm p} \tilde{\mathbf{v}}_{\rm p}} \mathbf{g}$ is the power received at the PT during the third stage, $\frac{1-\alpha}{2} T$ is the duration of the secondary transmissions, and $\eta$ is the power conversion efficiency.

The primary destination combines the received transmissions from both the PT and the secondary relays using the MRC technique. Thus, the combined signal at the primary destination is
 \begin{equation} \small
\small \begin{split} \small
y_{\rm pd}=\omega_1 \left(h_{\rm p,{\rm pd}} x+ w_2\right)+ \omega_2\left({\mathbf{g}^*}^\dagger \boldsymbol{h} x+{\mathbf{g}^*}^\dagger \tilde{\mathbf{v}} +w_3\right),
     \end{split}
\end{equation}
where $\omega_1$ and $\omega_2$ are the combining gains whose values are
\begin{equation} \small
\omega^*_1= \frac{h_{\rm p,{\rm pd}}^\dagger}{\kappa}, \  \omega^*_2=  \frac{\boldsymbol{h}^\dagger \mathbf{g}^*}{\kappa+{\mathbf{g}^*}^\dagger \boldsymbol{R}_{\tilde{\mathbf{v}}\tilde{\mathbf{v}}} \mathbf{g}^*}
\end{equation}


The rate of the PT in bits/sec/Hz is given by
\begin{equation} \small
\mathcal{R}_{\rm p}=\frac{1-\alpha}{2} \log_2\left( 1+\frac{P_{\rm p} \theta_{\rm p,pd}}{\kappa}+\frac{|{\mathbf{g}^*}^\dagger \boldsymbol{h}|^2}{\kappa+{\mathbf{g}^*}^\dagger \boldsymbol{R}_{\tilde{\mathbf{v}}\tilde{\mathbf{v}}} \mathbf{g}^*} \right).
\end{equation}

%
\vspace{-0.25cm}
\section{Simulation Results}\label{numerical}
\vspace{-0.1cm}
In this section, we provide simulation results to compare the achievable throughput of our PSA schemes. The system parameters used to generate the figures are: $4000$ time slots, $N=50$ secondary nodes,
$E_{\rm p}=50$ microJ, $W=1$ MHz, $T=1$ msec, $\eta=0.8$, $\kappa=0.01$ microWatts/Hz, $P_{\rm s}=0.1$ microWatts/Hz, and $K_R=5$ relays. We assume that all channel coefficients are distributed as circularly-symmetric Gaussian random variables with zero means and unit variances.

\begin{figure}
	\centering
		  \includegraphics[width=1\columnwidth]{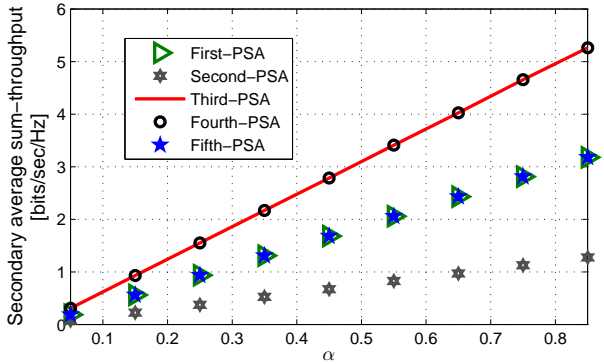}
	\caption{Average secondary sum-throughput in bits/sec/Hz versus $\alpha$.}
	\label{fig1}
\vspace{-0.5cm}
\end{figure}

Fig. \ref{fig1} shows the average secondary sum-throughput in bits/sec/Hz versus $\alpha$ for our PSA schemes. For the proposed schemes with beamformers, we assume that all nodes participate in powering the PT except the transmitter-receiver pair selected for data transmission. Fig. \ref{fig1} demonstrates that the secondary average sum-throughput increases with increasing $\alpha$ since increasing $\alpha$ increases the transmission times of the STs and, hence, increases their achievable rates. The third-PSA and fourth-PSA schemes have the same throughput which outperforms the other proposed schemes. This is because these schemes involve all the STs in data transmissions where the time is divided equally among the STs without any interference. Since these schemes enable the use of all channels between the STs and the PT, they achieve higher secondary throughput than the achieved throughput when using the maximum link-gain selection criterion for data transmission and PT powering. In addition, the first-PSA and fifth-PSA schemes have the second highest average secondary sum-throughput. The second PSA has the lowest secondary throughput due to the presence of interference among the STs which significantly degrades the achievable throughput. As an example, the achieved average sum-throughput is almost \emph{half} the achieved throughput by the first-PSA and fifth-PSA schemes.
\vspace{-0.0cm}

\begin{figure}
	\centering
		  \includegraphics[width=1\columnwidth]{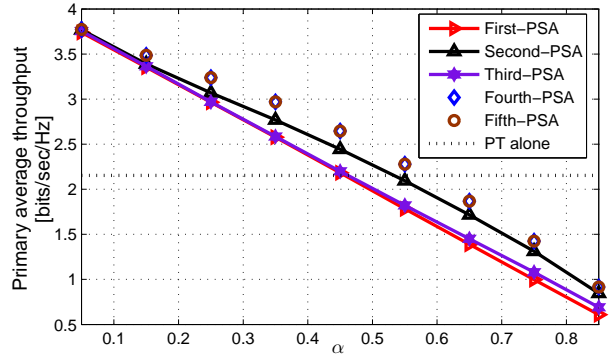}
	\caption{Average primary throughput in bits/sec/Hz versus $\alpha$.}
	\label{fig2}
\vspace{-0.45cm}
\end{figure}

Fig. \ref{fig2} shows that the primary throughput is monotonically decreasing with the parameter $\alpha$ since the time available for primary data transmission decreases with increasing $\alpha$. The case of PT alone, which represents the case when there is no cooperation between the primary and secondary systems, is plotted to show the gain of cooperation between both systems. As shown in the figure, when $\alpha=0.05$, all the proposed schemes achieve the same primary throughput due to the fact that the energy harvested at the PT from the secondary transmissions during the first stage is insignificant since $\alpha$ is very small. Accordingly, the throughput gain of the proposed schemes relative to the case of PT alone is due to relaying and the energy harvested during the third stage. The throughput gain is almost \emph{twice} the case when the PT is alone. When $\alpha$ exceeds $0.55$, all our proposed schemes provide lower primary throughput than the case when the PT is alone. However, we emphasize that the cooperation is still beneficial for the PT because its cooperation with the secondary nodes involves AF relaying with the goal of MSE minimization. Hence, the achieved probability of symbol error in the case of cooperation is lower than that achieved when the PT is alone. Note that the slope of the degradation of the primary throughput curves with $\alpha$ depends on the proposed scheme. The fourth-PSA and fifth-PSA schemes achieve the highest primary throughput among the proposed schemes. This is because the fourth-PSA and the fifth-PSA schemes use beamforming to power the PT which increases the received energy at the PT. The second PSA, which has the lowest secondary throughput in Fig. \ref{fig1}, achieves the second highest primary throughput. This is because the PT powering techniques under this scheme involve simultaneous data transmissions during the released time duration for the secondary nodes. Since the average power per secondary data transmission is $P_{\rm s}$, which is the average power used for any secondary activity to help the PT, the power received at the PT in this case is higher than the other schemes. The third-PSA scheme is slightly better than the first-PSA scheme in terms of the primary throughput. The fourth-PSA and fifth-PSA schemes can achieve up to $25 \%$ primary throughput gain relative to the third-PSA and fourth-PSA schemes.
\vspace{-0.15cm}
\section{Conclusions}\label{conclusions}
\vspace{-0.1cm}
In this paper, we proposed several SWIPT schemes for cooperative cognitive radio networks. In addition, we proposed a sparse relay-selection scheme that minimizes the MSE of the received primary data at the primary destination. Furthermore, we showed that using cooperative distributed beamforming increases the energy harvested at the PT and hence the achievable primary throughput. As shown in our numerical results, the beamforming-based PT-powering schemes can achieve maximum primary throughput gains up to $25 \%$ relative to the other proposed schemes. Our results demonstrated that the best secondary access schemes from a secondary throughput viewpoint are the ones which selects the highest link gain between the secondary links for data transmission. On the other hand, the secondary access schemes which use beamforming are the best powering schemes for the PT and thus achieve higher primary throughput. Moreover, for small $\alpha$, where $\alpha$ represents the time released for the secondary transmission and PT powering, our results showed that the power transfer during the first stage is insignificant and hence all access schemes achieve almost the same primary throughput. The case of small $\alpha$ demonstrates the gain of relaying and the power transferred during the third stage and shows that cooperation can \emph{double} the primary throughput gain compared to the case when the PT is alone.

\bibliographystyle{IEEEtran}
\bibliography{IEEEabrv,term_bib}
\vspace{-0.2cm}
\balance
\end{document}